\documentclass[pra,aps,onecolumn,10pt,showpacs,groupedaddress,superscriptaddress,floatfix,notitlepage]{revtex4-2}
\usepackage{amsmath,amsfonts,amssymb,graphics,graphicx,epsfig,color,times, braket}
\usepackage[utf8x]{inputenc}
\usepackage{color}
\usepackage{bbm, dsfont}
\usepackage{subfigure}
\usepackage{hyperref} 
\usepackage{mathrsfs}
\usepackage{verbatim} 
\usepackage{float}
\begin{document}

\title{Fast and High Excitation Transport in Waveguide Quantum Electrodynamics}

\author{Ya-Tang Yu}
\email{yatan1018@gmail.com}
\affiliation{Institute of Atomic and Molecular Sciences, Academia Sinica, Taipei 10617, Taiwan}

\author{I Gusti Ngurah Yudi Handayana}
\affiliation{Molecular Science and Technology Program, Taiwan International Graduate Program, Academia Sinica, Taiwan}
\affiliation{Department of Physics, National Central University, Taoyuan City 320317, Taiwan}
\affiliation{Institute of Atomic and Molecular Sciences, Academia Sinica, Taipei 10617, Taiwan}

\author{Wei Chen}
\affiliation{Department of Physics and Center for Theoretical Physics, National Taiwan University, Taipei 10617, Taiwan}
\affiliation{Center for Quantum Science and Engineering, National Taiwan University, Taipei 10617, Taiwan}
\affiliation{Institute of Atomic and Molecular Sciences, Academia Sinica, Taipei 10617, Taiwan}

\author{H. H. Jen}
\email{sappyjen@gmail.com}
\affiliation{Institute of Atomic and Molecular Sciences, Academia Sinica, Taipei 10617, Taiwan}
\affiliation{Molecular Science and Technology Program, Taiwan International Graduate Program, Academia Sinica, Taiwan}
\affiliation{Physics Division, National Center for Theoretical Sciences, Taipei 10617, Taiwan}

\date{\today}
\renewcommand{\r}{\mathbf{r}}
\newcommand{\f}{\mathbf{f}}
\renewcommand{\k}{\mathbf{k}}
\def\p{\mathbf{p}}
\def\q{\mathbf{q}}
\def\bea{\begin{eqnarray}}
\def\eea{\end{eqnarray}}
\def\ba{\begin{array}}
\def\ea{\end{array}}
\def\bdm{\begin{displaymath}}
\def\edm{\end{displaymath}}
\def\red{\color{red}}
\pacs{}
\begin{abstract}
Waveguide quantum electrodynamics (wQED) with underlying collective and long-range atom-atom interactions has led to many distinct dynamical phenomena, including modified collective radiations and intriguing quantum correlations. It stands out as a unique platform to illustrate correlated photon transport, as well as to promise applications in quantum information processing. Here we manifest a fast and high atomic excitation transport by employing two separated chirally-coupled atomic arrays. This enhanced waveguide-mediated transport of excitations emerges due to the dominance of few subradiant right eigenstates that are spectrally isolated and spatially localized in the system's dynamics. Contrary to the instinct of applying the cascaded systems with unidirectional couplings to expedite direct and high excitation transport, the optimal system configurations in open wQED systems demand slight or finite nonreciprocal decay channels to facilitate energy transport by exploiting waveguide-mediated couplings. We also investigate the effect of the couplings' directionality and the scaling of atom number on the transport properties. Our results showcase the wide applicability in wQED platforms and provide insights into quantum engineering and quantum information applications.
\end{abstract}
\maketitle
\section{Introduction}

Efficient energy transport in quantum systems is of fundamental interest and holds strong technological promise for quantum-enhanced applications \cite{Moreno-Cardoner2019,Moreno-Cardoner2022}. In the context of cold-atom platforms, facilitated energy transport plays a crucial role in applications such as quantum energy storage and conversion, offering promising routes toward the development of future energy-preserving devices. These capabilities rely on a key feature of atom-light interacting systems, the photon-mediated atom-atom interaction \cite{Jen2025}, which give rises to collective superradiant and subradiant channels, enables controllable enhancement or suppression of energy flow, and provides a versatile mechanism for engineering efficient and robust transport dynamics.

A delicate configuration of atoms in free space has been designed to facilitate an efficient transport of energy \cite{Needham2019,Moreno-Cardoner2019,Moreno-Cardoner2022} to envision light-harvesting applications \cite{Holzinger2024}. This is highly related to correlated photon transport that can further be enhanced by waveguide-mediated couplings \cite{Shen2007, Shen2007_2, Fan2010, Liao2010, Shi2011, Pletyukhov2012, Xu2015, Mahmoodian2018, Jones2020, Sheremet2023, Song2018, Iversen2021, Iversen2022, Shi2025}, where photon-photon correlations can be modified as bunching or antibunching due to collectively enhanced nonlinearity \cite{Prasad2020}. A counterintuitive slowdown dynamics of photon switching is also predicted for large atom arrays  \cite{Poddubny2024} due to quantum Zeno effect \cite{Koshino2005}. These intriguing transport features with underlying nontrivial and nonclassical correlations showcase the rich opportunities in waveguide quantum electrodynamics (wQED) \cite{Jen2020_disorder, Zhong2020, Jen2021_bound, Fayard2021, Jen2022_correlation, Wu2024, Yudi2025, Pichler2015, Lodahl2017, Sheremet2023} for quantum state engineering and quantum information applications \cite{Chien2024, Goswami2025_efficient}.

Recent studies focus on waveguide-coupled energy transport among two distant atomic ensembles \cite{Fasser2024} and two separate collectively driven quantum emitters \cite{vanDiepen2025}, which reveal the essential roles of the superradiant sectors that expedite the emission or absorption of light and the subradiant sectors that can be selectively populated as quantum storage \cite{Ferioli2021}. This promises deterministic generation of photonic entangled states \cite{Tudela2015, Bigorda2025}, a resource as the building blocks for quantum communication \cite{Duan2001} and quantum network \cite{Kimble2008}. With the flexible positioning of atoms assisted by optical tweezer arrays, tunable directionality of couplings \cite{Mitsch2014}, and infinite-range photon-mediated dipole-dipole interactions \cite{Solano2017, Tudela2024, Jen2025}, the ultimate capability of efficient excitation transport in wQED setups is yet to be unveiled. This raises the question of how wQED platforms can be fine-tuned to exhibit fast and high energy transport, while also elucidating the essential underlying mechanisms.
 
 \begin{figure}[t]
 	\centering
 	\includegraphics[width=0.45\textwidth]{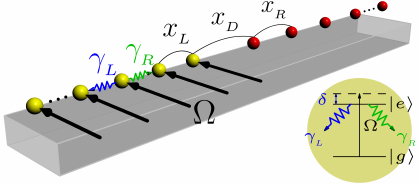}
 	\caption{Schematic of the system. A two-level atomic array coupled to a waveguide with asymmetric decay rates, $\gamma_L \ne \gamma_R$, corresponding to the left- and right-propagating modes. The atoms are divided into two spatially separated groups with a distance $x_D$ between them, each with interparticle spaces  $x_L$ and $x_R$, respectively. A coherent drive with a Rabi frequency $\Omega$ is applied only to the atoms in the left group and the detuning of atoms in the rotating frame of the driving frequency is $\delta$. The coupling to the guided modes induces infinite-range or all-to-all dipole-dipole interactions among atoms, including coherent exchange and collective dissipation. This setting provides a minimal model for studying excitation transport between driven and undriven atomic arrays in the low-energetic excitation regime, where the spatial arrangement plays a key role in shaping the steady-state population distributions.}\label{Fig1}
\end{figure}
 
In this article, we demonstrate an efficient transport of atomic excitations among two separated emitter arrays via waveguide-coupled dipole-dipole interactions [See Fig. \ref{Fig1}]. In the low-energy excitation sectors, we identify the optimal configurations that lead to such outperformed excitation transport. They are associated with the dominance of few subradiant right eigenstates that are spectrally isolated and localized in the destination array. In contrast to the conventional wisdom using the unidirectional couplings to expedite one-way excitation transport \cite{Gardiner1993, Carmichael1993, Stannigel2012, Downing2020, Guimond2020}, the enhanced excitation transport emerges with the assistance of nonreciprocal couplings. This facilitated transport of excitations can be attributed to the intricate interplay between the directionality of couplings and the photon-mediated dipole-dipole interactions among all atoms under driven-dissipative conditions.

\section{System model}

To explore excitation transport, we consider a one-dimensional (1D) array of two-level atoms coupled to a chiral waveguide supporting asymmetric decay channels, as shown in Fig. \ref{Fig1}. The atomic array is divided into two spatially separated groups, $N_L$ atoms in the left group and $N_R$ atoms in the right group, for a total of $N=N_L+N_R$ atoms. The left and right groups are arranged with uniform interatomic spacings $x_L$ and $x_R$, respectively, and the distance between the last atom in the left group and the first atom in the right group is $x_D$. We apply a weak coherent drive with a Rabi frequency $\Omega$ to the atoms in the left group, and the detuning of the atoms is $\delta$ in the rotating frame. This spatially asymmetric configuration allows us to explore excitation transport from driven to undriven regions. We note that all the numerical results in this work are under the resonant condition where the detuning $\delta=0$, which means the drive is resonant with the emitters.

By applying the Born–Markov approximation and the rotating-wave approximation, the dynamics of the atoms' density matrix $\rho$ is governed by the master equation for a generic driven-dissipative 1D chirally-coupled atomic array ($\hbar=1$) \cite{Pichler2015, Sheremet2023},
\bea
\label{eq: master}
\dot\rho=
-i[H_S+H_L+H_R,\rho]
+\mathcal{L}_L[\rho]+\mathcal{L}_R[\rho], 
\eea
where the system Hamiltonian $H_S=
\sum_{\mu=1}^{N}-\delta\sigma^\dagger_\mu\sigma_\mu+
\sum_{\mu=1}^{N_L}\Omega (\sigma_\mu+\sigma^\dagger_\mu)$ is composed of the detunings of atoms and a uniform external drive to the atoms in the left group, the coherent interaction and dissipative Lindblad terms are $H_{L(R)}=-i\frac{\gamma_{L(R)}}{2}
\sum_{\mu<(>)\nu}^N
(e^{ik_s|x_\mu-x_\nu|}
\sigma^\dagger_\mu\sigma_\nu-e^{-ik_s|x_\mu-x_\nu|}
\sigma^\dagger_\nu\sigma_\mu)$ and $\mathcal{L}_{L(R)}[\rho]=\gamma_{L(R)}
\sum_{\mu,\nu=1}^N
e^{\mp ik_s(x_\mu-x_\nu)}
[\sigma_\nu\rho\sigma^\dagger_\mu
-\{\sigma^\dagger_\mu\sigma_\nu,\rho\}/2]$, respectively. All terms above are expressed in the rotating frame of the driving frequency, and $k_s$ is the corresponding effective wave vector of the guided modes. For atom $\mu$ at site $x_\mu$, the raising and lowering dipole operators are $\sigma_\mu^\dagger\equiv\ket{e_\mu}\bra{g_\mu}$ and $\sigma_\mu=(\sigma_\mu^\dagger)^\dagger$ with the ground state $|g_\mu\rangle$ and the excited state $\ket{e_\mu}$. The decay channels are denoted as left- and right-propagating modes $\gamma_L$ and $\gamma_R$, respectively, with $\gamma_L\neq\gamma_R$ in general. We introduce the directionality factor $D\equiv(\gamma_R-\gamma_L)/\gamma$ \cite{Mitsch2014} and the total decay rate $\gamma\equiv\gamma_L+\gamma_R$ for the following discussions.

To investigate the low-energy dynamics, one can reduce the master equation to an effective non-Hermitian Hamiltonian and use the Schrödinger equation to describe the system’s evolution \cite{Ashida2020, Torres2014, Kumlin2018, Jen2020_steady}. See Appendix A for the derivation of the effective Hamiltonian and a discussion on the domain of validity of this approach. In the weak driving limit $\Omega \ll \gamma$ (we set $\Omega/\gamma = 10^{-3}$ for all numerical simulations in this work), the system remains predominantly in the ground state $\ket{G} \equiv \ket{g}^{\otimes N}$. We can therefore express the total system state as a perturbative expansion $\ket{\Psi_{\text{tot}}(t)} \approx \ket{G} + \ket{\Psi(t)}$, where $\ket{\Psi(t)}$ represents the dynamical single-excitation component. We denote the $N$ singly-excited bare states as $\ket{\psi_\mu}\equiv\ket{e_\mu}\ket{g}^{\otimes (N-1)}$ for each site. The system dynamics can thus be described by the singly-excited state function $\ket{\Psi(t)}=\sum_{\mu=1}^N p_\mu(t)\ket{\psi_\mu}$ under the condition that $\sum_{\mu=1}^N |p_\mu|^2\ll 1$. By substituting the total expansion $\ket{\Psi_{\text{tot}}(t)}$ into the Schrödinger equation and projecting onto the single-excitation subspace, we obtain the evolution of the individual probability amplitudes:
\bea
\label{eq: singleEQ}
\dot{p}_\mu(t)=M_{\mu\nu}p_\nu(t)-i\tilde\Omega_\mu.
\eea
Here, the effective interaction matrix $M$ accounts for the dissipative coupling and detunings, while the source term $-i\tilde{\Omega}_\mu$ arises specifically from the driving laser coupling the highly-populated ground state $\ket{G}$ to the excited state $\ket{\psi_\mu}$. Given that the system remains close to the ground state ($\ket{\Psi_{\text{tot}}(t)} \approx \ket{G}$), the driving interaction can be approximated as $\sum_{\mu=1}^{N_L}\Omega (\sigma_\mu+\sigma^\dagger_\mu)\ket{\Psi_{\text{tot}}(t)} \approx \sum_{\mu=1}^{N_L}\Omega \sigma^\dagger_\mu\ket{G}$. The driving vector is thus defined as $\tilde\Omega/\Omega\equiv (1,\dots,1,0,\dots,0)$, where the elements are finite when $\mu\in[1, N_L]$, and the effective interaction matrix $M$ can be expressed as
\bea
M_{\mu\nu}
=\begin{cases}
-\gamma_Le^{ik_s|x_\mu-x_\nu|}, & \mu<\nu\\
i\delta-\frac{\gamma_L+\gamma_R}{2}, & \mu=\nu\\
-\gamma_Re^{ik_s|x_\mu-x_\nu|}, & \mu>\nu\\
\end{cases}
.
\eea

By diagonalizing the matrix $M$, we can further derive the system dynamics in terms of its right eigenstates \cite{Ashida2020, Torres2014}. For the system initially at the ground state $\ket{g}^{\otimes N}$, which implies an initial condition $p_\mu(0)=0$,
\bea
\label{eq: solution}
\ket{\Psi(t)} = -\Omega \sum_{n=0}^{N-1}
\frac{\Delta_n}{\mathcal{E}_n}
\left(1 - e^{-i\mathcal{E}_n t}\right) \ket{\phi_n^R},
\eea
where $\ket{\phi_n^R}$ is the normalized right eigenstate with the complex eigenenergy $\mathcal{E}_n=\omega_n-i\gamma_n$, $\omega_n$ and $\gamma_n$ the corresponding energy shift and decay rate, respectively. Furthermore, by projecting this state back onto the bare state basis, we obtain the probability amplitude as 
\bea
	\label{eq: Coefficient}
	p_\mu(t)
	=-\Omega\sum_{n=0}^{N-1}
	\frac{\Delta_n}{\mathcal{E}_n}
	(1-e^{-i\mathcal{E}_nt})
	\braket{\psi_\mu|\phi_n^R}.
\eea
For those eigenmodes decaying slower than natural decay ($\gamma_n < \gamma$), they are known as subradiant modes. $\Delta_n=\bra{\phi_n^L}(\sum_{\mu=1}^{N_L}\ket{\psi_\mu})$ is the projected component of the driving subspace on the left eigenstate $\ket{\phi_n^L}$. The above expression shows two essential ingredients that determine the dynamics of the system and its steady-state behaviors: $\mathcal{E}_n^{-1}$ represents the respective weight on each right eigenstate, and the sustained steady states at infinite time are thus dominated by several most subradiant modes with small magnitude of complex energy $|\mathcal{E}_n|$. We note that Eq. (\ref{eq: solution}) does not apply to the reciprocal cases ($D=0$) at $x_L,x_R,x_D=m\pi/k_s$ with integers $m$ and the unidirectional case ($D=1$). The former corresponds to a Bragg-spaced array \cite{Sheremet2023}, leading to a breakdown of using only the low-energetic sectors, with divergences in state populations due to the presence of decoherence-free states, while the latter becomes non-diagonalizable due to the appearance of exceptional points \cite{Ashida2020, Moiseyev2011} where eigenvalues and eigenstates coalesce. The latter case, however, can be solved by integral forms \cite{Jen2020_collective} for the cascaded systems or by numerical simulations directly.

\section{Excitation transport and characteristic time}

Next, we analyze the excitation transport properties by calculating the excitation distributions of the quasi-steady state and the associated characteristic time to reach it. While transport can occur transiently, the excitation may subsequently return to the left array during the evolution, making such transfer effects fleeting. We therefore focus on the quasi-steady state, as it provides a stable and experimentally reliable measure of sustained transport, robust against minor fluctuations in measurement timing. We utilize the normalized state functions to extract an excitation distribution profile, which remains independent of the driving amplitude in the low-energy regime, depends only on the system parameters, and provides useful information on the quasi-steady states before reaching the steady states at $t\rightarrow \infty$. The normalized state function can be obtained as 
\bea
\label{eq: profile}
\ket{\tilde\Psi(t)}&&=\frac{1}{\sqrt{\sum_{\nu=1}^N|\braket{\psi_\nu|\Psi(t)}|^2}}\sum_{\mu=1}^N\braket{\psi_\mu|\Psi(t)}\ket{\psi_\mu}, \\
&&\equiv\sum_{\mu=1}^N\tilde p_\mu(t)\ket{\psi_\mu}, \nonumber
\eea
which reflects the relative distributions of excitations among all sites. We then define the excitation transport parameter $T_p(t)$ as the difference in excitation populations $\tilde P_\mu\equiv|\tilde p_\mu|^2$ between the right and the left groups of atoms in the steady-state profile \cite{Jen2019}, 
\bea
\label{eq: Tp}
T_p=
\sum_{\mu=N_L+1}^{N}\tilde{P}_\mu(\infty)
-\sum_{\mu=1}^{N_L}\tilde{P}_\mu(\infty). 
\eea
A complete excitation transport indicates $T_p=1$, while $T_p=-1$ indicates the opposite where all excitation populations remain in the left and driven group.   

\begin{figure}[h]
	\centering
	\includegraphics[width=0.5\textwidth]{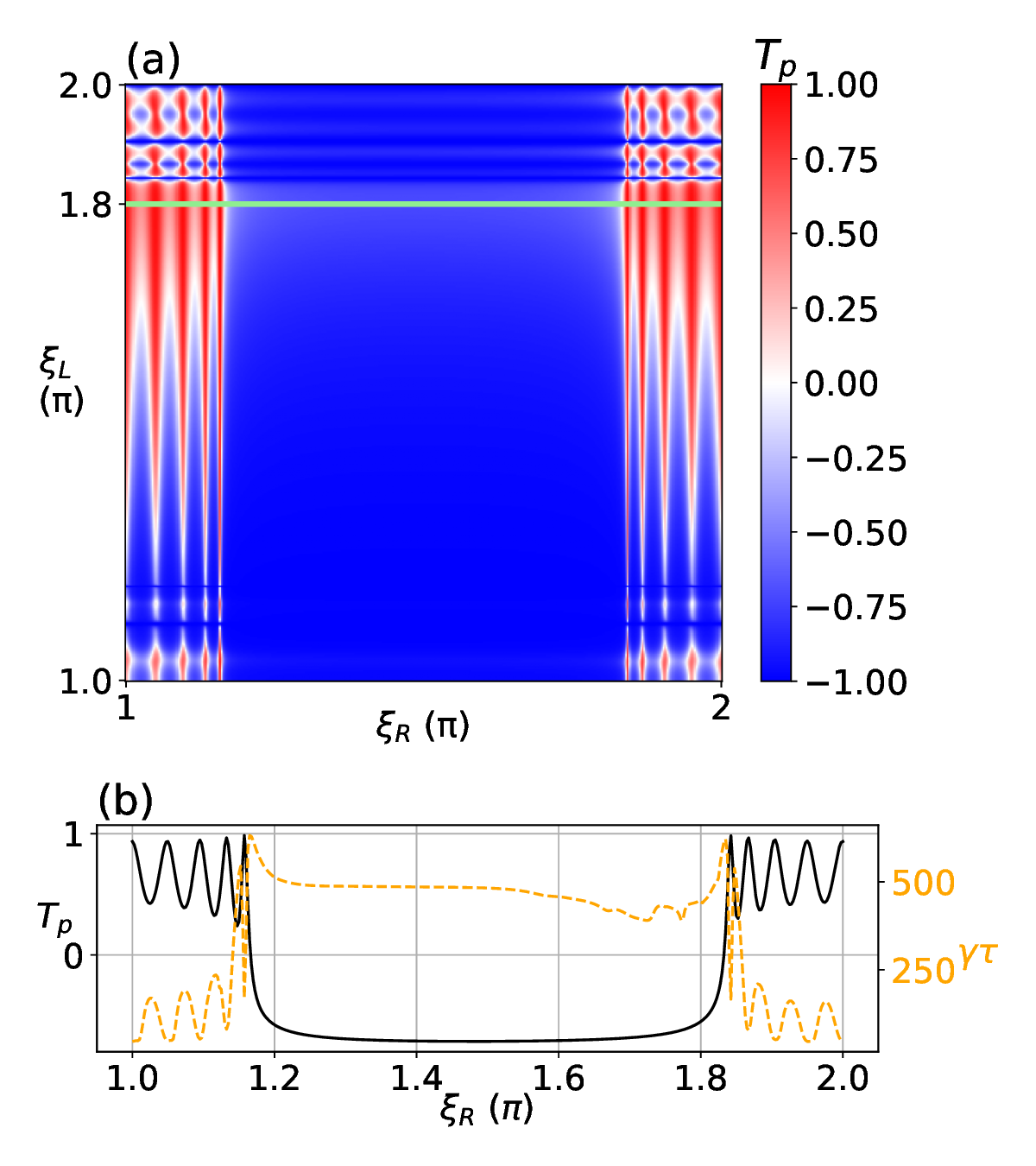}
	\caption{Excitation transport $T_p$ and characteristic time $\tau$. (a) The dependence of $T_p$ on $\xi_L$ and $\xi_R$ at $\xi_D=1.5\pi$ and $D=0.5$. A total of $N=20$ atoms with $N_L=N_R$. (b) A cross-sectional cut in (a) with $\xi_L=1.8\pi$ (the light green line), where $T_p$ (black solid line) is plotted with the associated $\tau$ (orange dashed line) versus $\xi_R$. We observe that relatively high $T_p$ values coincide with relatively short $\tau$ values, indicating the optimal conditions of $\xi_R$ for efficient excitation transport from the driven group to the undriven one.}\label{Fig.2}
\end{figure}

The timescale to reach the steady-state $T_p$ is essential and also informative to quantify the efficiency of the excitation transport. In practice, it is often sufficient for the system to reach a quasi-steady state, when the excitation profile becomes nearly time-independent. To characterize this timescale, a characteristic time $\tau$ can well be approximated as the average time weighted by the infidelity between the instantaneous and steady-state excitation profiles, 
\bea
\label{eq: Tau}
\tau=\frac{\int_0^\infty t (1-|\braket{\tilde{\Psi}(\infty)|\tilde{\Psi}(t)}|^2)dt}{\int_0^\infty  (1-|\braket{\tilde{\Psi}(\infty)|\tilde{\Psi}(t)}|^2)dt}. 
\eea
The use of state infidelity bounds the weighted distributions at long time, which guarantees the convergence of the denominator in Eq. (\ref{eq: Tau}). 

In Fig. \ref{Fig.2}, we show the excitation transport properties $T_p$ and the characteristic time $\tau$ versus the dimensionless interparticle spacings $\xi_L\equiv k_sx_L$ and $\xi_R\equiv k_sx_R$, at certain $\xi_D\equiv k_sx_D$ and the directionality $D$. We find that a relatively high $T_p$ coincides with a relatively short $\tau$. This manifests the parameter regimes to achieve both high and fast excitation transport from the driven (left) to the undriven (right) group. Notably, the high excitation transport emerges near $\xi_R\approx \pi$ and $2\pi$ with an interference pattern as seen in Fig. \ref{Fig.2}(b). This alternating appearance of optimal $T_p$'s can be extended as $D$ increases and with more fringes as $N_R$ increases, and similarly these increased fringes also emerge as $N_L$ increases along the cross sectional cut at a particular $\xi_R$, as shown in Appendix B. This reflects the influence of photon-mediated dipole-dipole interactions which exchange excitations among all atoms, leading to interference fringes. We note that a translation symmetry is sustained such that $T_p$ and $\tau$ remain intact when $\xi_{R, D}\leftrightarrow \xi_{R, D}+\pi$. 

\section{The right-localized right eigenstates}

\begin{figure*}[t] 
	\centering
	\includegraphics[width=0.9\textwidth]{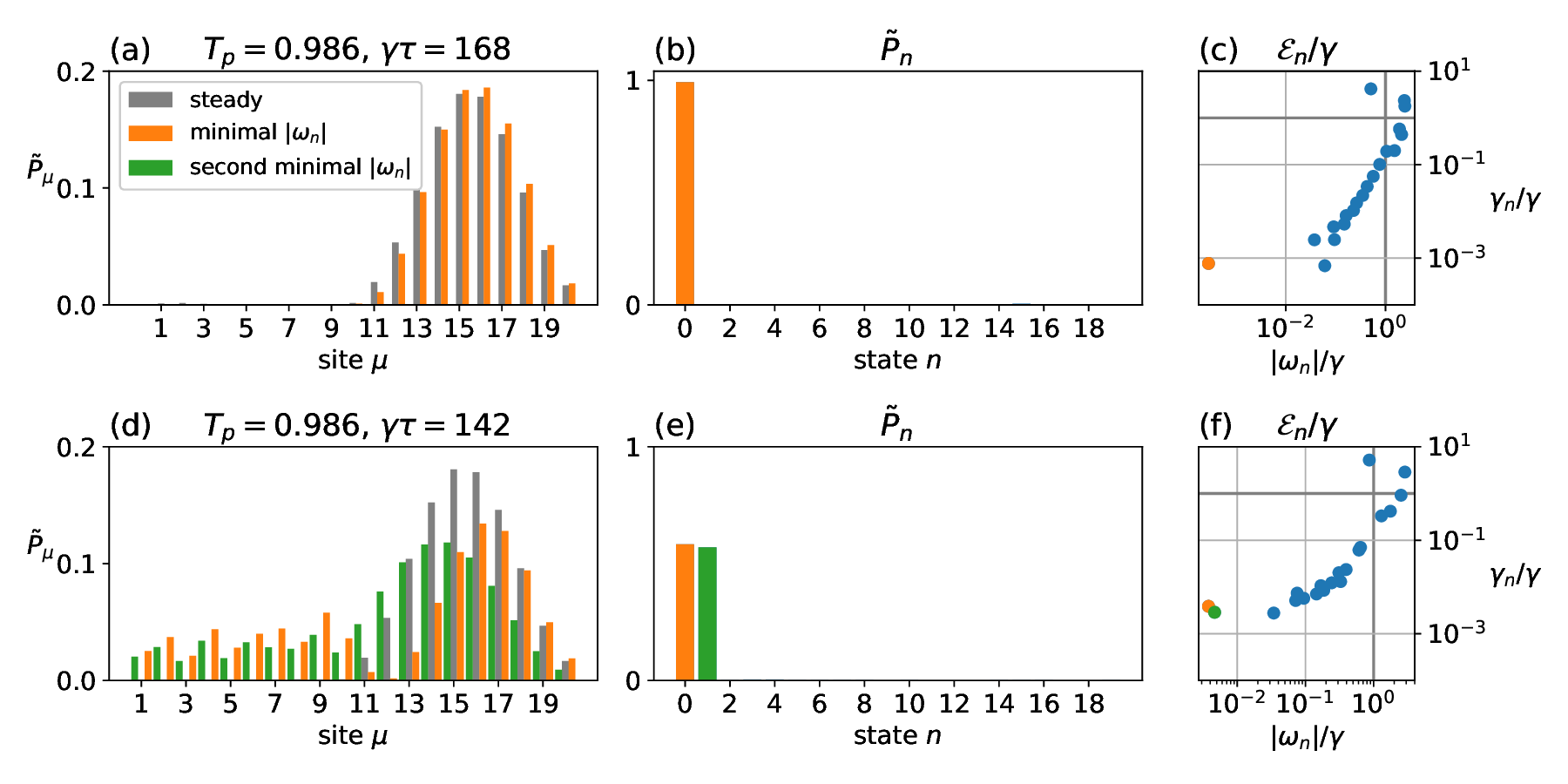}
	\caption{Characterization of the steady states with high excitation transport. The parameters are chosen as $N = 20$, $D = 0.5$ for (a, b, c) $\xi_L = 1.8\pi$, $\xi_D = 1.5\pi$, $\xi_R = 1.158\pi$, and (d, e, f) $\xi_L = 1.96\pi$, $\xi_D = \xi_R=1.158\pi$, where we observe an optimal $T_p$ and $\tau$. (a) Site-resolved excitation profile $\tilde{P}_\mu$ of the steady state (gray bars) compared with that of the right eigenstate (orange bars) with the smallest eigenenergy magnitude, $|\mathcal{E}_0|$. This also corresponds to the mode with the minimal frequency shift $\omega_0$ of the interaction matrix $M$ as shown in (c), where most of the excitations are localized in the right group. (b) Spectrum of the steady-state population decomposed onto the normalized right eigenspace of $M$, ordered by ascending values of $|\omega_n|$. The dominant contribution results from the mode whose weight $\mathcal{E}_n^{-1}$ significantly exceeds all other components. (c) Complex eigenenergies $\mathcal{E}_n=\omega_n-i\gamma_n$ plotted in logarithmic scales in the complex plan. The orange dot marks the mode with minimal eigenenergy shift $|\omega_0|$, which lies closest to the origin compared to the rest of the modes (blue dots), showing a long-lived and spectrally isolated behavior. Similarly, (d) excitation profile $\tilde P_\mu$ and two right eigenstates (orange and green bars) with their spectral decomposition in (e) and complex energies in (f).}\label{Fig.3} 
\end{figure*}

Focusing on the parameter regimes where fast and high excitation transport emerges, we look into the steady-state excitation profile and the associated right eigenstates. In Fig. \ref{Fig.3}, we identify two kinds of cases which are responsible for high $T_p\approx 1$. One corresponds to a single right-localized right eigenstate of the non-Hermitian interaction matrix $M$ as illustrated in Figs. \ref{Fig.3}(a) and  \ref{Fig.3}(b). Another one corresponds to two right eigenstates which jointly contribute to a right-localized population profile in Figs. \ref{Fig.3}(d) and  \ref{Fig.3}(e). They demonstrate significant populations localized in the right atomic group in the steady state and for relatively fast transport due to the spectrally-isolated right eigenmodes. When the steady-state components consist primarily of a single or few spectrally-close eigenstates, the system quickly evolves into that configuration without interference from the other modes. In contrast, when the steady-state excitation components are spread across multiple right eigenstates, the system’s evolution involves a dynamic exchange among several subradiant modes. This interplay delays the convergence to the steady state. Such exchange arises from the non-Hermitian nature of the system, where right eigenstates are typically non-orthogonal.

Notably, the essential feature for high and fast excitation transport lies at few most subradiant modes that are spectrally-isolated as shown in Fig. \ref{Fig.3}(c), where the minimal $|\mathcal{E}_0|\ll|\mathcal{E}_n|, \forall n\ne0$, and in Fig. \ref{Fig.3}(f), where the minimal and second minimal $|\mathcal{E}_0|,|\mathcal{E}_1|\ll|\mathcal{E}_n|,\forall n\ne 0,1$. These states generally correspond to the minimal frequency shifts $|\omega_0|,|\omega_1|$, not necessarily to the most subradiant modes that $\gamma_0,\gamma_1\ge\min(\gamma_n)$. A relatively small value of $|\mathcal{E}_n|$ implies both slow decay and slow oscillation, rendering the dominant mode the most dynamically stable state in the system. Once the excitation population flows into this state, it is unlikely to escape, effectively being `trapped' on the right and undriven part of the system. This highlights that spectral isolation, rather than the minimal decay alone, is key to long-lived excitation confinement.   

For $T_p\approx -1$ or $0$ as the lowest transport or almost balanced transport scenarios, respectively, similar single or double spectrally-isolated modes can emerge, as shown in Figs. \ref{Fig.S6} and \ref{Fig.S7} in Appendix C. Therefore, a fast and high excitation transport always corresponds to few subradiant eigenmodes which are spectrally-isolated from the rest of the modes, speeding up the evolution due to their dynamical stability. However, the opposite does not hold since it could involve more than few right eigenmodes that inhibit the transport. 

\begin{figure*}[t]
	\centering
	\includegraphics[width=0.9\textwidth]{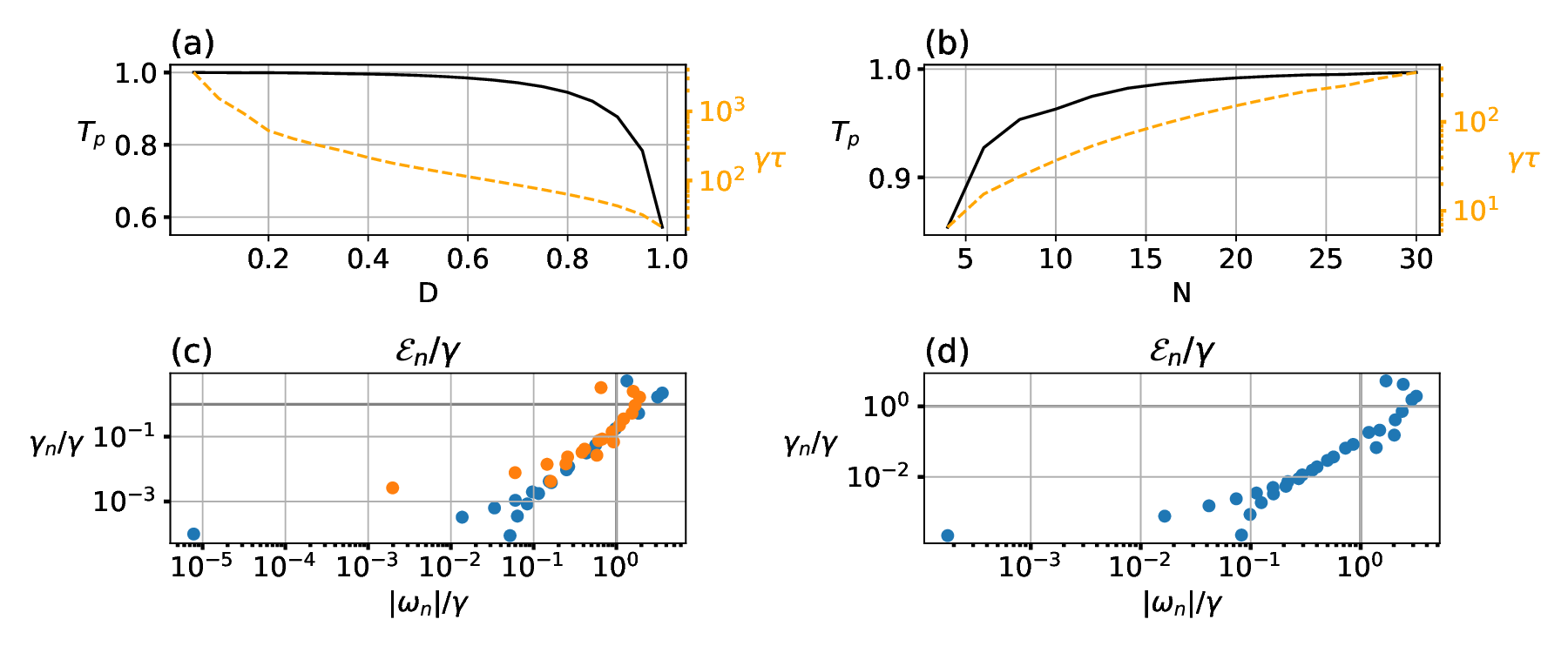} 
	\caption{The effect of directionality factor $D$ and scaling of atom number $N$. With the atoms equally partitioned ($N_L=N_R$), we find the optimal interatomic spacing configuration that indicates fast and high excitation transport and plot the associated $T_p$ and $\tau$ versus $D$ in (a) for $N=20$ and versus even total atom number $N$ in (b) for $D=0.5$. (c) The corresponding complex eigenenergy plots for $D=0.2$ (Blue dots) and $0.8$ (orange dots) in (a). (d) The corresponding spectrally-isolated modes for $N=30$ in (b).}\label{Fig.4}
\end{figure*}

\section{Effect of $D$ and scaling of $N$}

As for the effect of the directionality in the waveguide-mediated couplings and the scaling of total number of atoms, we extract the optimal interatomic spacings configuration that indicates fast and high excitation transport and plot the corresponding $T_p$ and characteristic time $\tau$ in Fig. \ref{Fig.4}. In Fig. \ref{Fig.4}(a), we find a decreasing $T_p$ as $D$ increases, but with a decreasing $\tau$, indicating a trade-off between the transport quantity and its speed. The suppression of excitation transport with increasing directionality, reaching a minimum under unidirectional coupling, reflects the essential role of a significant degree of reciprocal couplings that mediate the spin-exchange interactions among all atoms. This further enables a high excitation transport and coincides with a spectrally-isolated subradiant eigenmode shown in Fig. \ref{Fig.3}, Figs. \ref{Fig.4}(c), and \ref{Fig.4}(d), which would not generally be allowed in noninteracting quantum emitters. The utility of large $D$, on the other hand, accelerates the speed to reach the steady-state profile as expected, but with a steep drop in $T_p$ as $D$ approaches $1$.  

As $N$ increases, the optimal $T_p$ found in Fig. \ref{Fig.4}(b) also increases, and the corresponding $\tau$ increases as well, suggesting another trade-off between the transport quantity and its speed. Large system size promotes excitation transport since waveguide-mediated couplings render the subradiant modes increasingly dominant in the system dynamics. This collective and enhanced transport performance gradually saturates as $N$ further increases, but at a price of low efficiency due to large $\tau$. This can be attributed to the longer timescale (smaller $\gamma_n$) for the dominant and spectrally-separated subradiant mode in Fig. \ref{Fig.4}(d). 

\section{Discussion and Conclusion}

For experimental feasibility, here we assume that the atoms are perfectly coupled to the waveguide. While realistic systems may exhibit imperfect coupling, our results remain robust in strongly coupled platforms, as discussed in Appendix D, such as  superconducting qubits \cite{Chang2018, Kim2021} and quantum dots \cite{vanDiepen2025}. For other platforms—including atomic arrays \cite{Goban2015, Douglas2015, Holzinger2022} coupled to photonic crystal waveguides, atom-nanofiber systems with atoms trapped in optical lattices \cite{Corzo2019}, and atom-waveguide systems employing optical tweezers \cite{Samutpraphoot2020, Dordevic2021}-the coupling efficiency is typically lower, and special designs (e.g., introducing two or more waveguides \cite{Kritika2024}) may be required to enhance the coupling strength. The proposed pumping scheme can be realized via controlled side excitations \cite{Mitsch2014}, and state characterizations can be achieved through individual site-resolved measurements. Therefore, provided that atom-waveguide coupling is sufficiently strong, our results are experimentally promising and should be readily implemented in the state-of-the-art platforms \cite{Sheremet2023}.

We uncover a fast and high excitation transport in wQED platforms by employing two separate chirally-coupled atomic arrays. Applying low-energetic excitation analysis reveals the crucial parameter regimes that give rise to fast and high excitation transport. To summarize the optimized configurations identified in this work, the efficiency of excitation transport is predominantly determined by the interparticle spacing $\xi_{\rm R}$ particularly when chosen near $\pi$ or $2\pi$ in the regime of low directionality ($D<0.5$). For larger directionality, excitation transport is generally suppressed for the entire parameter space spanned by the three interparticle spacings, making this regime less favorable. Finally, while the dependence on $\xi_{\rm L}$ and $\xi_{\rm D}$ is weaker, the transport efficiency can be further fine-tuned by adjusting these parameters. This enhanced waveguide-coupled transport of excitations indicates the dominance of few spectrally-isolated right eigenstates, extremely localized in the undriven region of the atoms. This localization enables a large population imbalance in favor of the undriven group, while the spectral isolation of the dominant eigenstate ensures rapid convergence to the steady state profile. Our findings reveal that both spatial localization and spectral structure of the underlying non-Hermitian modes play key roles in optimizing transport. This insight could serve as a guiding principle in engineering fast and directional excitation transport in quantum optical systems, with potential applications in quantum information processing and quantum communication.

\section{Acknowledgments}

We acknowledge support from the National Science and Technology Council (NSTC), Taiwan, under the Grants No. 112-2112-M-001-079-MY3 and No. NSTC-114-2119-M-001-005, and from Academia Sinica under Grant AS-CDA-113-M04. We are also grateful for support from TG 1.2 of NCTS at NTU.  

\section{Data Availability Statement}

The data and the codes to support the results of this article are available \cite{Zenodo}.  
\appendix

\section{Derivation of the Effective Hamiltonian and its Domain of Application}
The derivation begins with the master equation (Eq. 1 in the main text):
\begin{align} 
\label{eq: master}
\dot\rho=
-i[H_S+H_L+H_R,\rho]
+\mathcal{L}_L[\rho]+\mathcal{L}_R[\rho], 
\end{align}
where the system Hamiltonian is given by
\begin{align}
\label{eq: H_S}
H_S=
\sum_{\mu=1}^{N}-\delta\sigma^\dagger_\mu\sigma_\mu+
\sum_{\mu=1}^{N_L}\Omega (\sigma_\mu+\sigma^\dagger_\mu).
\end{align}
The coherent interaction is described by $H_L(R)$ and the dissipation by the Lindblad terms $\mathcal{L}_{L(R)}$:
\begin{align}
\label{eq: H_L(R)}
H_{L(R)}=-i\frac{\gamma_{L(R)}}{2}
\sum_{\mu<(>)\nu}^N
(e^{ik_s|x_\mu-x_\nu|}
\sigma^\dagger_\mu\sigma_\nu-e^{-ik_s|x_\mu-x_\nu|}
\sigma^\dagger_\nu\sigma_\mu),
\end{align}
and
\begin{align}
\label{eq: L_L(R)}
\mathcal{L}_{L(R)}[\rho]=\gamma_{L(R)}
\sum_{\mu,\nu=1}^N
e^{\mp ik_s(x_\mu-x_\nu)}
[\sigma_\nu\rho\sigma^\dagger_\mu
-\{\sigma^\dagger_\mu\sigma_\nu,\rho\}/2].
\end{align}
All of these terms are expressed in the rotating frame with respect to the driving frequency, and $k_s$ is the corresponding effective wave vector in this frame.

Our goal is to rearrange the master equation into the form:
\begin{align} 
\label{eq: master-eff}
\dot\rho=
-i(H_\text{eff}\rho-\rho H^\dagger_\text{eff})
+\sum_{\mu,\nu=1}^N
[e^{ik_s(x_\mu-x_\nu)}+e^{-ik_s(x_\mu-x_\nu)}]\sigma_\nu\rho\sigma^\dagger_\mu,
\end{align}
where the effective non-Hermitian Hamiltonian $H_\text{eff}$ combines all terms excluding the quantum jumps. The effective Hamiltonian can be decomposed into a Hermitian part ($H_\text{H}$) and a purely non-Hermitian part ($H_\text{nH}$) as $H_\text{eff}=H_\text{H}+H_\text{nH}$. The Hermitian part is $H_\text{H}=H_S+H_L+H_R$, and the purely non-Hermitian part arises from the spin-conserved parts of the Lindblad terms $\mathcal{L}_{L(R)}[\rho]$. By applying the following mathematical trick $\{A,B\}=-i\{iA,B\}=-i[(iA)B-B(iA^\dagger)^\dagger]$ for two arbitrary operators $A$ and $B$, we can rewrite the Lindblad terms as follows:
\begin{align} 
\label{eq: Lindblad}
\mathcal{L}_{L(R)}[\rho]=i\gamma_{L(R)}
\sum_{\mu,\nu=1}^N
e^{\mp ik_s(x_\mu-x_\nu)}
[(i\sigma^\dagger_\mu\sigma_\nu)\rho-\rho(i\sigma^\dagger_\nu\sigma_\mu)^\dagger]
+\gamma_{L(R)}
\sum_{\mu,\nu=1}^N
e^{\mp ik_s(x_\mu-x_\nu)}
\sigma_\nu\rho\sigma^\dagger_\mu.
\end{align}

Now, by combining the first term on the right-hand side of Eq. (\ref{eq: Lindblad}) with the Hermitian terms, we obtain the effective Hamiltonian:
\begin{align} 
\label{eq: effective}
H_\text{eff}=H_S
-i\left[
\gamma_L\sum_{\mu<\nu}^N
e^{ik_s|x_\mu-x_\nu|}
\sigma^\dagger_\mu\sigma_\nu
+\gamma_R\sum_{\mu>\nu}^N
e^{ik_s|x_\mu-x_\nu|}
\sigma^\dagger_\mu\sigma_\nu
+\frac{\gamma_L+\gamma_R}{2}\sum_{\mu=\nu}^N
\sigma^\dagger_\mu\sigma_\mu\right].
\end{align}

Next, we explain why the system's dynamics in the low-energy regime can be described using only the effective non-Hermitian Hamiltonian.

The evolution of an arbitrary observable $\hat O$ is governed by the Heisenberg equation:
\begin{align} 
\label{eq: Heisenberg}
\frac{d\hat O}{dt}=
i[H_S+H_L+H_R,\hat O]
+\tilde{\mathcal{L}}_L[\hat O]+\tilde{\mathcal{L}}_R[\hat O],
\end{align}
where the adjoint dissipative Lindblad terms are given by:
\begin{align} 
\label{eq: adj-Lindblad}
\tilde{\mathcal{L}}_{L(R)}[\hat O]=\gamma_{L(R)}
\sum_{\mu,\nu=1}^N
e^{\mp ik_s(x_\mu-x_\nu)}
[\sigma^\dagger_\mu\hat O\sigma_\nu
-\{\sigma^\dagger_\mu\sigma_\nu,\hat O\}/2],
\end{align}
The evolution of the expectation value of this observable is then given by $\frac{d}{dt}\braket{\hat O}=\braket{\frac{d\hat O}{dt}}$.

The observables we are interested in are the atomic excitation populations at each site, $\ket{e}_\alpha\bra{e}_\alpha=\sigma^\dagger_\alpha\sigma_\alpha$. For these observables, the quantum jump terms in Eq. (\ref{eq: adj-Lindblad}) take the form $\sigma^\dagger_\mu\sigma^\dagger_\alpha\sigma_\alpha\sigma_\nu$. 
In the low-energy regime, the dynamics are restricted to the ground state and the singly-excited subspace. Within this subspace, any operator containing the double raising operators (like $\sigma^\dagger_\alpha\sigma^\dagger_\mu$) or  double lowering operators (like $\sigma_\mu\sigma_\nu$) has a zero expectation value. Consequently, the expectation value of the quantum jump contribution vanishes:
\begin{align} 
\label{eq: meanValueJump}\braket{\sigma^\dagger_\alpha\sigma^\dagger_\mu\sigma_\mu\sigma_\nu}=0,\ 
\forall \alpha,\mu,\nu\in[1,N]
\end{align}
Therefore, the quantum jumps do not affect the evolution of the excitation populations. This justifies describing the system's dynamics solely with the effective non-Hermitian Hamiltonian and reducing the master equation to a Schrödinger equation for the state function $\ket{\Psi_\text{tot}(t)}$:
\begin{align} 
\label{eq: Schrodinger}
\frac{d}{dt}\ket{\Psi_\text{tot}(t)}
=-iH_\text{eff}\ket{\Psi_\text{tot}(t)}.
\end{align}

\section{Phase Diagram Pattern Affected by the Parameters}

In this work, we have six parameters to explore: the directionality of propagation ($D$), the atom numbers in the left and right groups ($N_L$, $N_R$), and three dimensionless interatomic spacings ($\xi_L$, $\xi_D$, and $\xi_R$). To analyze how these parameters influence excitation transport $T_p$, we examine their dependencies one by one.

\begin{figure}[h]
	\centering
	\includegraphics[width=0.77\textwidth]{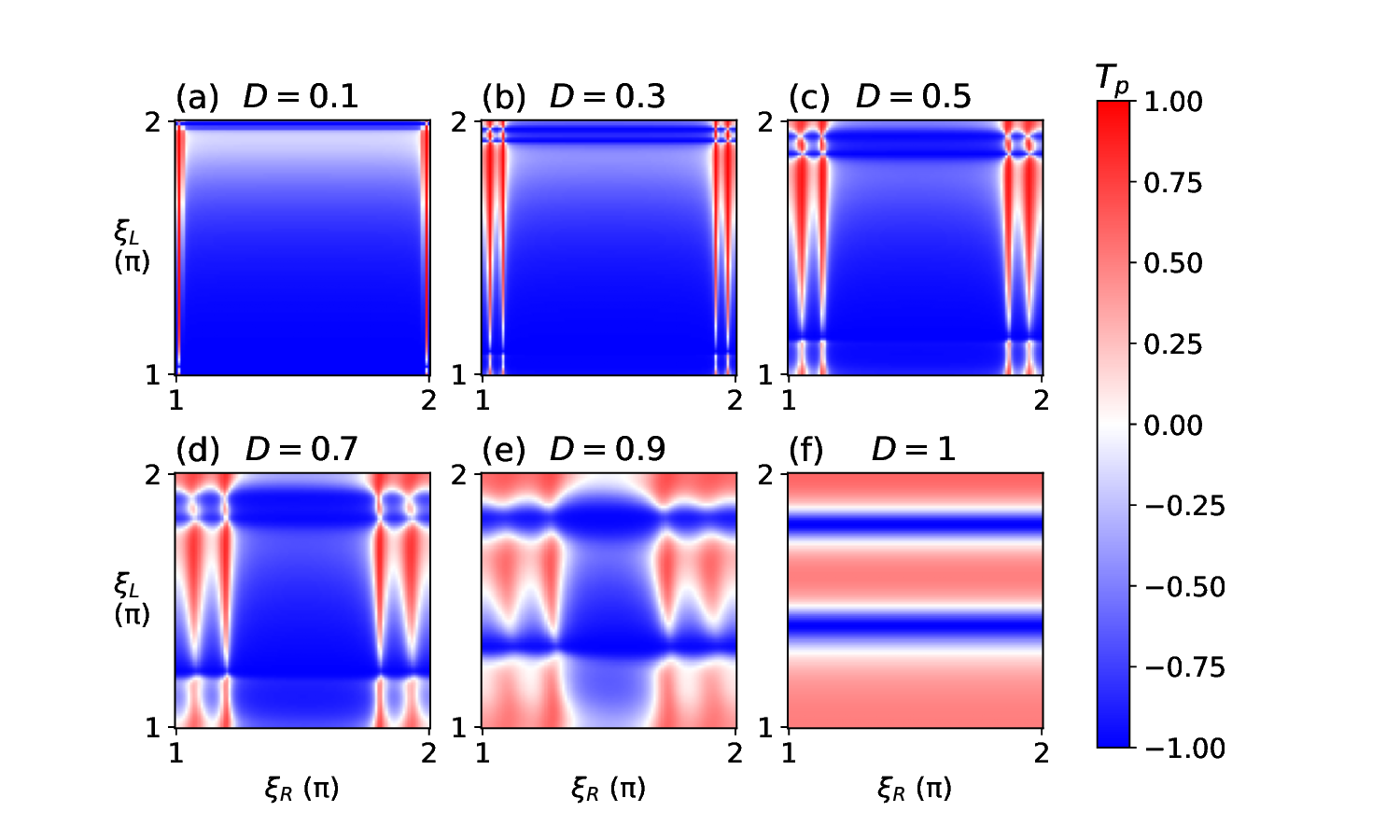}
	\caption{Phase diagrams of $T_p$ with respect to $\xi_L$ and $\xi_R$ in the range $[\pi, 2\pi]$, at fixed $\xi_D = 1.5\pi$ and $N_L = N_R = 5$, for different values of directionality $D$ (indicated at the top of each panel). As $D$ increases toward the unidirectional limit, the high-$T_p$ regions become broader but the values of $T_p$ decrease.
	}\label{Fig.S1}
\end{figure}

\subsection{Effect of Directionality}
Fig. \ref{Fig.S1} shows phase diagrams of $T_p$ with respect to $\xi_L$ and $\xi_R$ in the region $[\pi, 2\pi]$, for various values of directionality $D$, at fixed $\xi_D = 1.5\pi$ and atom numbers $N_L = N_R = 5$.
In Fig. \ref{Fig.S1}(a), where $D = 0.1$ (close to the reciprocal limit), high-$T_p$ (indicated by red areas) regions are narrowly concentrated around $\xi_R = \pi$ and $2\pi$. As directionality increases from (b) to (e), the high-$T_p$ zones expand toward $\xi_R = 1.5\pi$, but the intensity of the transport (depth of red) diminishes. In the fully unidirectional case, Fig. \ref{Fig.S1}(f) with $D = 1$, the vertical fringes of high-$T_p$ regions disappear. This insensitivity to $\xi_R$ arises because the system enters a cascaded regime where atomic spacing $\xi_R$ imparts only site-dependent phase factors to the steady state amplitudes, which cancel out in the population calculation. These merge into a few horizontal bands, and the overall excitation transport becomes small across the entire parameter space. In conclusion, as $D$ approaches 1 (the unidirectional limit), the regions with relatively high excitation transport become broader, but the magnitudes in these regions decrease.

\begin{figure}[h]
	\centering
	\includegraphics[width=0.56\textwidth]{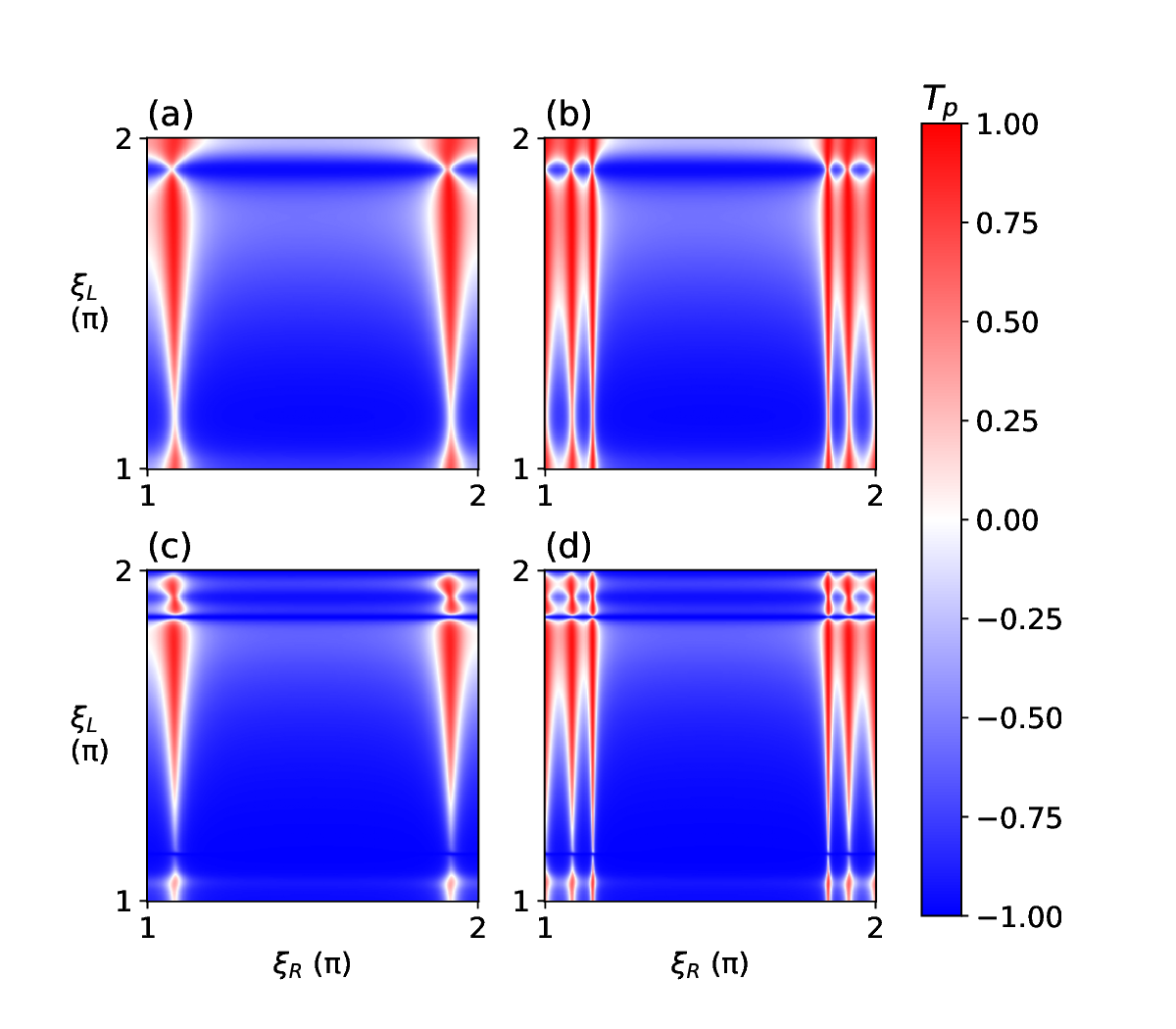}
	\caption{Phase diagrams of $T_p$ with respect to $\xi_L$ and $\xi_R$ in the region $[\pi, 2\pi]$, at fixed directionality $D = 0.5$ and $\xi_D = 1.5\pi$, for different combinations of $N_L$ and $N_R$:
		(a) $N_L=3$, $N_R=3$;
		(b) $N_L=3$, $N_R=6$;
		(c) $N_L=6$, $N_R=3$;
		(d) $N_L=6$, $N_R=6$.
		For (a) and (c), which share $N_R = 3$, high-$T_p$ regions align along two primary lines. For (b) and (d), with $N_R = 6$, the number of high-$T_p$ lines increases to six. These results confirm that the $\xi_R$-dependent structure of transport is governed by $N_R$, while $\xi_L$-dependence relates to $N_L$.
	}\label{Fig.S2}
\end{figure}
\begin{figure}[h]
	\centering
	\includegraphics[width=0.75\textwidth]{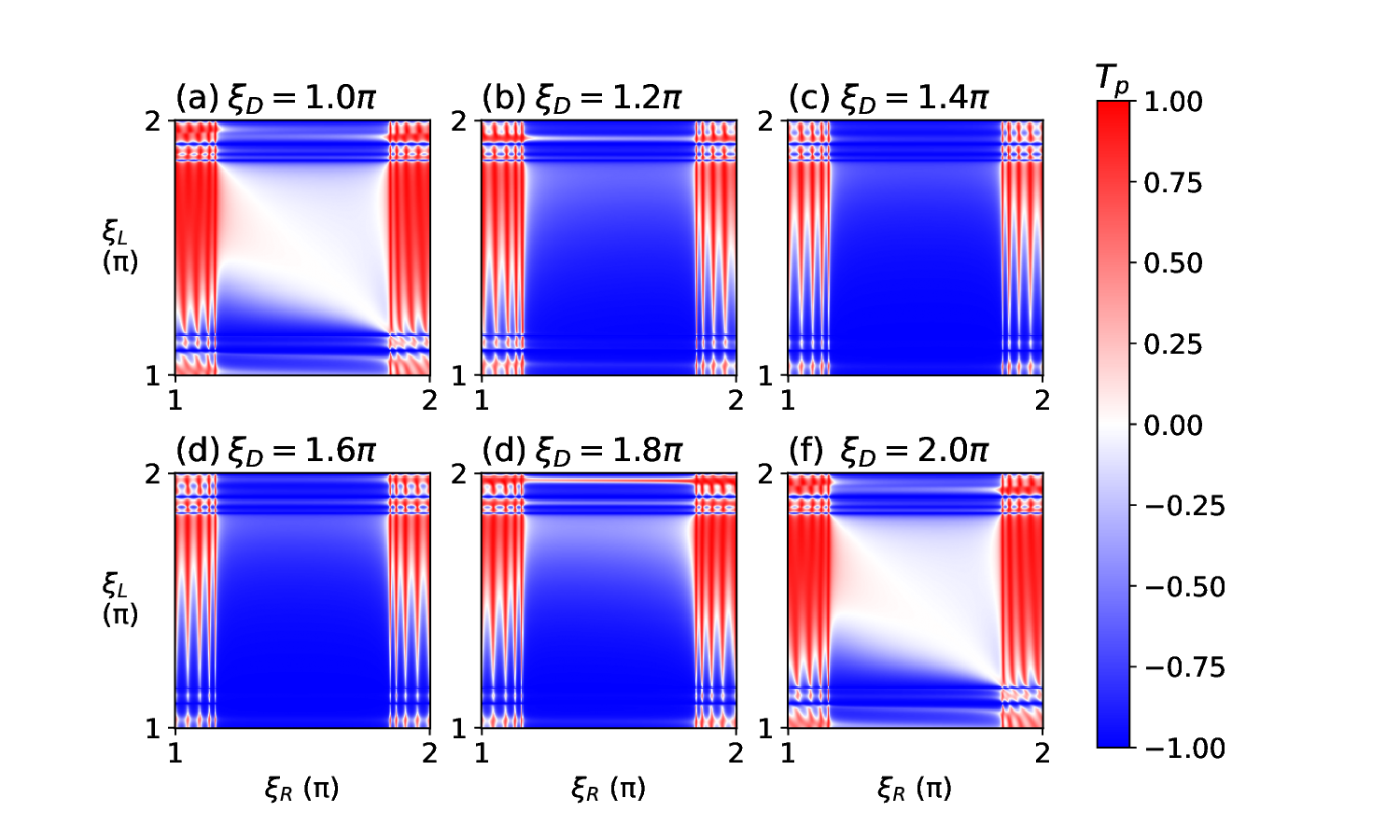}
	\caption{$T_p$ phase diagrams versus $\xi_L$ and $\xi_R$ in the region $[\pi,2\pi]$ at fixed $N_L=N_R=10$, $D=0.5$ for different $\xi_D$ (labeled at the top of each diagram). The high-$T_p$ areas are strongly correlated to $\xi_R$ (always located around the same lines).}\label{Fig.S3}
\end{figure}

\subsection{Effect of Atom Number}
From the $T_p$ phase diagrams for several combinations of $N_L$ and $N_R$, shown in Fig. \ref{Fig.S2}, we observe that regions of high transport (red areas) form around multiple distinct lines (the fringes) in the parameter space. As $N_R$ increases, the number of these fringes also increases. Notably, the number of high-$T_p$ fringes correlates with the number of atoms on the right side, observed as $2\times\lfloor N_R/2 \rfloor$. This scaling arises from the number of right-localized subradiant eigenmodes. Similarly, regions of low transport (blue areas) form horizontal fringes, and the number increases with $N_L$.

\subsection{The Roles of Interatomic Spacings}

To understand how the three interatomic spacings affect excitation transport, we plot the phase diagrams of $T_p$ with respect to $\xi_L$ and $\xi_R$ for several fixed $\xi_D$, under fixed directionality and atom numbers. As shown in Fig. \ref{Fig.S3}, regions of high $T_p$ consistently align with specific values of $\xi_R$, showing little relation to $\xi_L$ and $\xi_D$. This suggests that excitation transport is predominantly governed by $\xi_R$. Based on this observation, an effective strategy to optimize $T_p$ is to first identify favorable values of $\xi_R$, and then fine-tune $\xi_L$ and $\xi_D$ to further locate the optimum of enhanced excitation transport.

\section{Representative Steady-State Profiles for low and negative $T_p$}

Here we investigate some representative steady-state profiles at $T_p\approx -1$ and $0$, respectively, as a comparison to the high $T_p\approx 1$ discussed in the main text. 

\begin{figure}[h]
	\centering
	\includegraphics[width=0.8\textwidth]{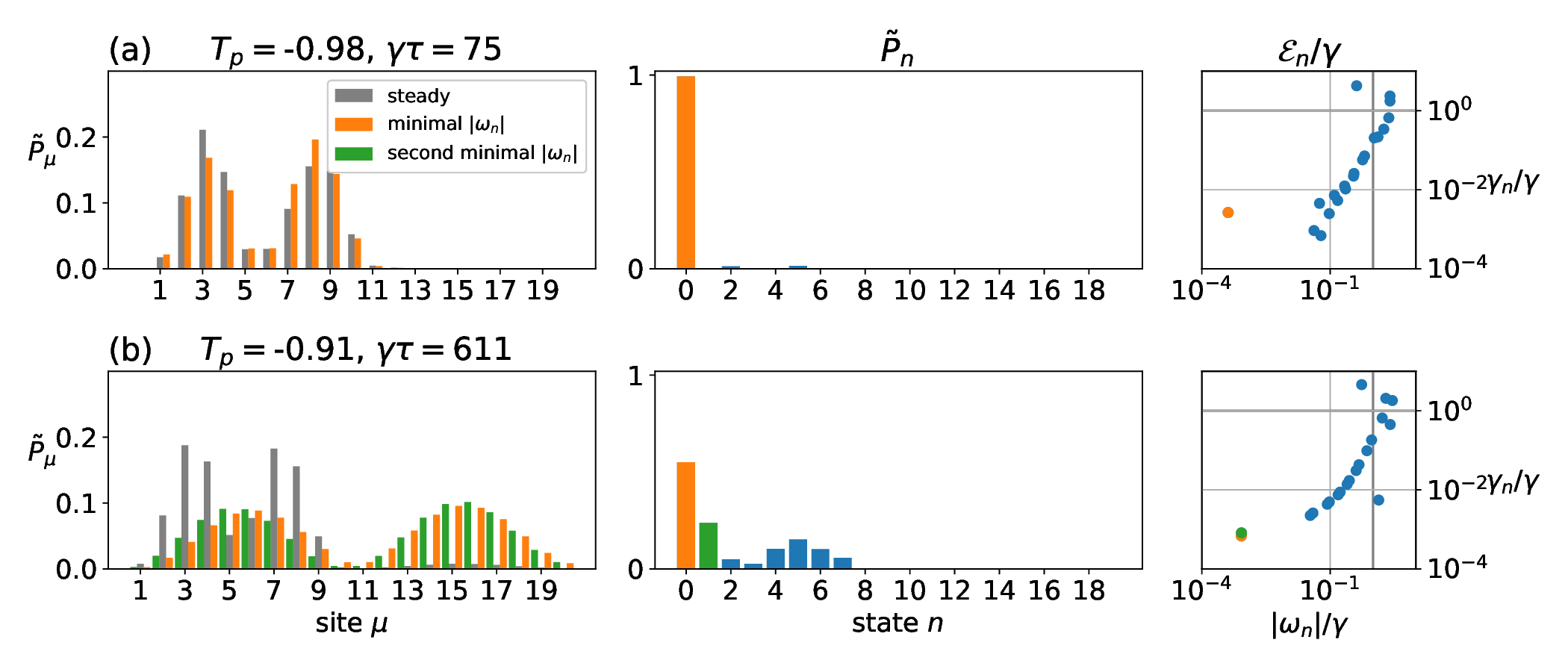}
	\caption{Steady-state profiles for two strong-storage ($T_p\approx-1$) parameter sets. All cases are simulated with $N_L=N_R=10$, and $D=0.5$, focusing on variations in interatomic spacings. Each case includes three subfigures arranged in three panels:
		Left panels show the site-resolved excitation profile $\tilde{P}_\mu$ of the steady-state profile (gray bars), compared with the minimal $|\omega_n|$ state (orange bars) and the second minimal $|\omega_n|$ state (green bars);
		Middle panels show the spectrum of the steady-state population decomposed into the normalized right eigenspace of $\mathcal{M}$, the right eigenstates are ordered by ascending values of $|\omega_n|$;
		Right panels display the complex eigenenergies $\mathcal{E}_n$ plotted on a logarithmic scale in the complex plane.  
		In Case (a), the minimal $|\omega_n|$ state is left-localized. Case (b) shows a profile composed of multiple right eigenstates, resulting in slower dynamics.
		Parameters used: 
		(a) $\xi_L = 1.133\pi$, $\xi_D = 1.5\pi$, $\xi_R = 1.8\pi$;  
		(b) $\xi_L = 1.158\pi$, $\xi_D = 1.5\pi$, $\xi_R = 1.158\pi$; 
	}\label{Fig.S6}
\end{figure}

Figure \ref{Fig.S6}(a,b) shows steady-state profiles with strong excitation storage on the left side ($T_p \approx -1$). In Case (a), the situation is similar to Fig. \ref{Fig.3}(a), except that the eigenmode with the minimal energy shift is left-localized, leading to dominant excitation retention in the left array. In Case (b), the steady-state profile is shaped by both the minimal and second minimal $|\omega_n|$ states, which are spectrally isolated from the rest. However, the characteristic time is longer than in Case (a) because the excitation continues to oscillate among several eigenmodes—ranging from the minimal $|\omega_n|$ to the seventh minimal—before reaching the steady-state profile.

\begin{figure*}[h]
	\centering
	\includegraphics[width=0.8\textwidth]{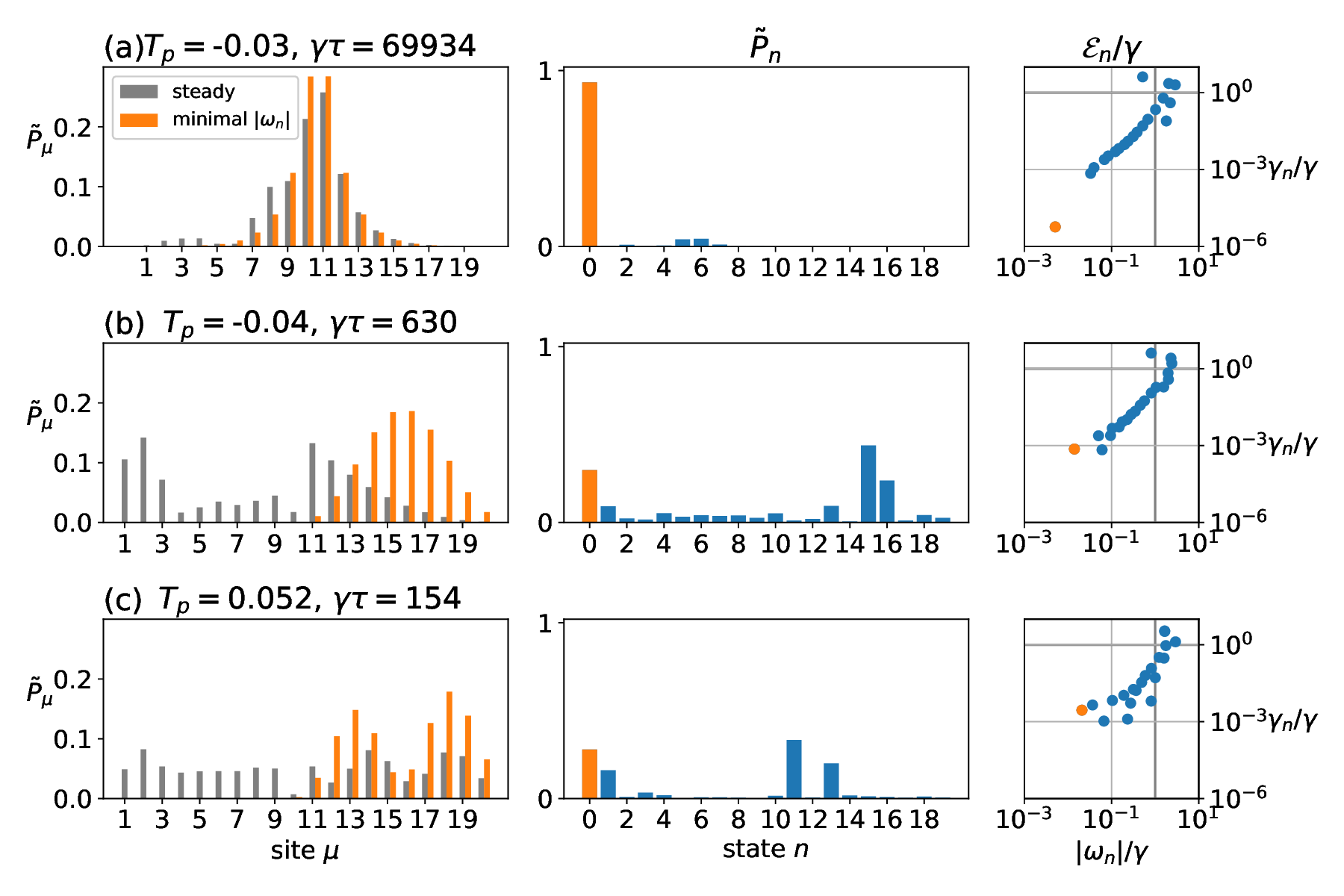}
	\caption{Steady-state profiles for parameter sets with no population imbalance ($T_p\approx0$) parameter sets. All cases are simulated with $N_L=N_R=10$, and $D=0.5$, focusing on variations in interatomic spacings. Each case includes three subfigures arranged in three panels:
		Left panels show the site-resolved excitation profile $\tilde{P}_\mu$ of the steady-state profile (gray bars), compared with the minimal $|\omega_n|$ state (orange bars) and the second minimal $|\omega_n|$ state (green bars);
		Middle panels show the spectrum of the steady-state population decomposed into the normalized right eigenspace of $\mathcal{M}$, the right eigenstates are ordered by ascending values of $|\omega_n|$;
		Right panels display the complex eigenenergies $\mathcal{E}_n$ plotted on a logarithmic scale in the complex plane.  
		Case (a) illustrates a center-localized profile. Cases (b,c) demonstrate the profiles that have no population imbalance and are not center-localized.
		Parameters used:  
		(a) $\xi_L = 1.18\pi$, $\xi_D = 1.1\pi$, $\xi_R = 1.18\pi$;  
		(b) $\xi_L = 1.8\pi$, $\xi_D = 1.5\pi$, $\xi_R = 1.167\pi$;
		(c) $\xi_L = 1.66\pi$, $\xi_D = 1.5\pi$, $\xi_R = 1.88\pi$.
	}\label{Fig.S7}
\end{figure*}

Figure \ref{Fig.S7} illustrate profiles with no population imbalance ($T_p \approx 0$). In Case (a), the steady-state profile is primarily composed of the minimal $|\omega_n|$ state, which is localized near the center of the array, neither left- nor right-localized. Unlike in Fig. \ref{Fig.3} (a), the minimal $|\omega_n|$ state does not fully dominate the dynamics here; other eigenmodes, notably the 5th and 6th, maintain non-negligible populations. This allows population exchange between states and results in a relatively long characteristic time $\tau$. In contrast, Cases (b) and (c) show weakly biased profiles that are not center-localized. In conclusion, spatially localized and spectrally isolated states play a crucial role in enabling both fast transport and efficient excitation storage.

\section{Effects of Waveguide Imperfections}

The preceding analysis assumed an idealized regime where waveguide-mediated interactions dominate the dynamics, and the atomic arrays are perfectly ordered with identical emitters. However, experimental implementations inevitably introduce imperfections. In this section, we assess the robustness of our protocol against three key realistic factors: (i) dissipative loss to non-guided modes, (ii) positional disorder within the arrays, and (iii) inhomogeneous broadening of the atomic transitions. We first address the effect of non-guided modes by introducing the coupling efficiency $\beta$ \cite{Sheremet2023}, and subsequently examine the impact of disorder.

\subsection{Effect of Non-Guided Modes}
We add two terms—the coherent interaction $H_\text{ng}$ from the nonguided effect and the corresponding dissipative Lindblad term $\mathcal{L}_\text{ng}$—into the Eq. (\ref{eq: master}) to model the dynamics of the imperfect waveguide system:

\begin{align} 
	\label{eq: Imperfect-master}
	\dot\rho=
	-i[H_S+H_L+H_R+H_\text{ng},\rho]
	+\mathcal{L}_L[\rho]+\mathcal{L}_R[\rho]+\mathcal{L}_\text{ng}[\rho].
\end{align}
Given that the free-space interaction between atoms decreases sharply with distance, and assuming that all interatomic spacings ($x_L$, $x_R$ and $x_D$) are much larger than the atomic emission wavelength ($x_L,x_R,x_D\gg\lambda=c/\omega$), the most significant effect from the nonguided modes is on-site dissipation. Thus, $H_\text{ng}$ can be neglected, and the dissipative Lindblad term can be approximated as:

\begin{align}
	\label{eq: L_ng}
	\mathcal{L}_\text{ng}[\rho]=
	\gamma_\text{ng}
	\sum_{\mu=1}^N
	[\sigma_\mu\rho\sigma^\dagger_\mu
	-\{\sigma^\dagger_\mu\sigma_\mu,\rho\}/2],
\end{align}
where $\gamma_\text{ng}$ is the decay rate into the nonguided modes. We introduce the coupling efficiency factor $\beta=\frac{\gamma}{\gamma+\gamma_\text{ng}}$ which shows the ratio between guided decay rate and nonguided decay rate.

\begin{figure*}[h]
	\centering
	\includegraphics[width=0.8\textwidth]{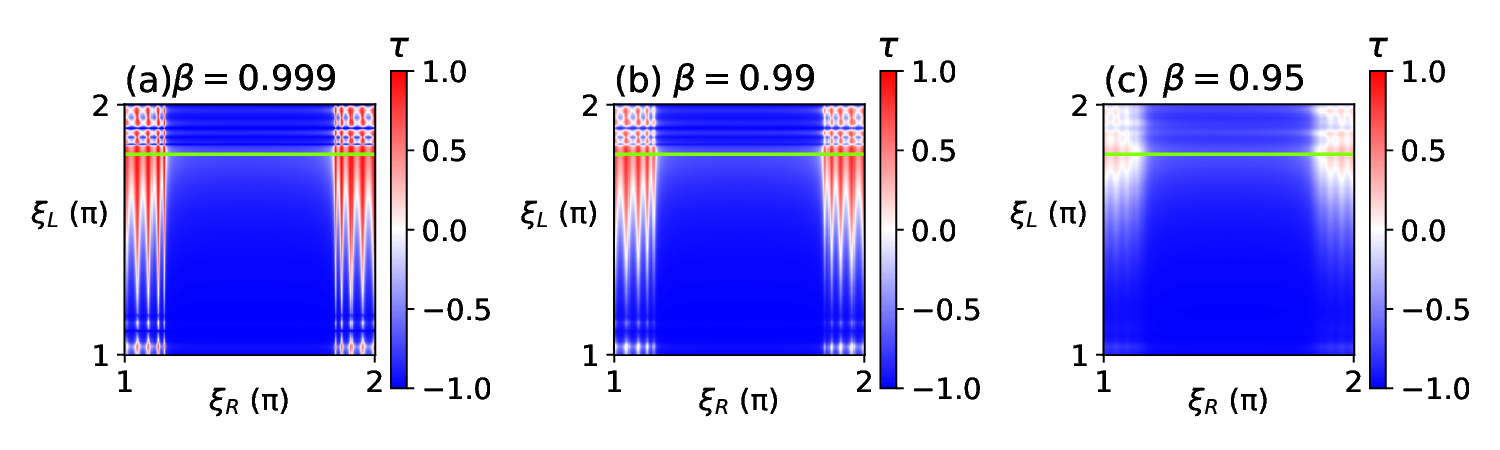}
	\caption{Phase diagrams of $T_p$ with respect to $\xi_L$ and $\xi_R$ for different $\beta$. One can see that the intensity of the fringes decreases with $\beta$. All cases are simulated with $N_L=N_R=10$, and $D=0.5$, with $\xi_D = 1.5\pi$. The x-axis and y-axis represent $\xi_R$ and $\xi_L$, respectively, both scaled in units of $\pi$. Fig. \ref{Fig.S9} is plotted on the cross-sectional cut (the light green line).
	}\label{Fig.S8}
\end{figure*}

\begin{figure*}[h]
	\centering
	\includegraphics[width=0.63\textwidth]{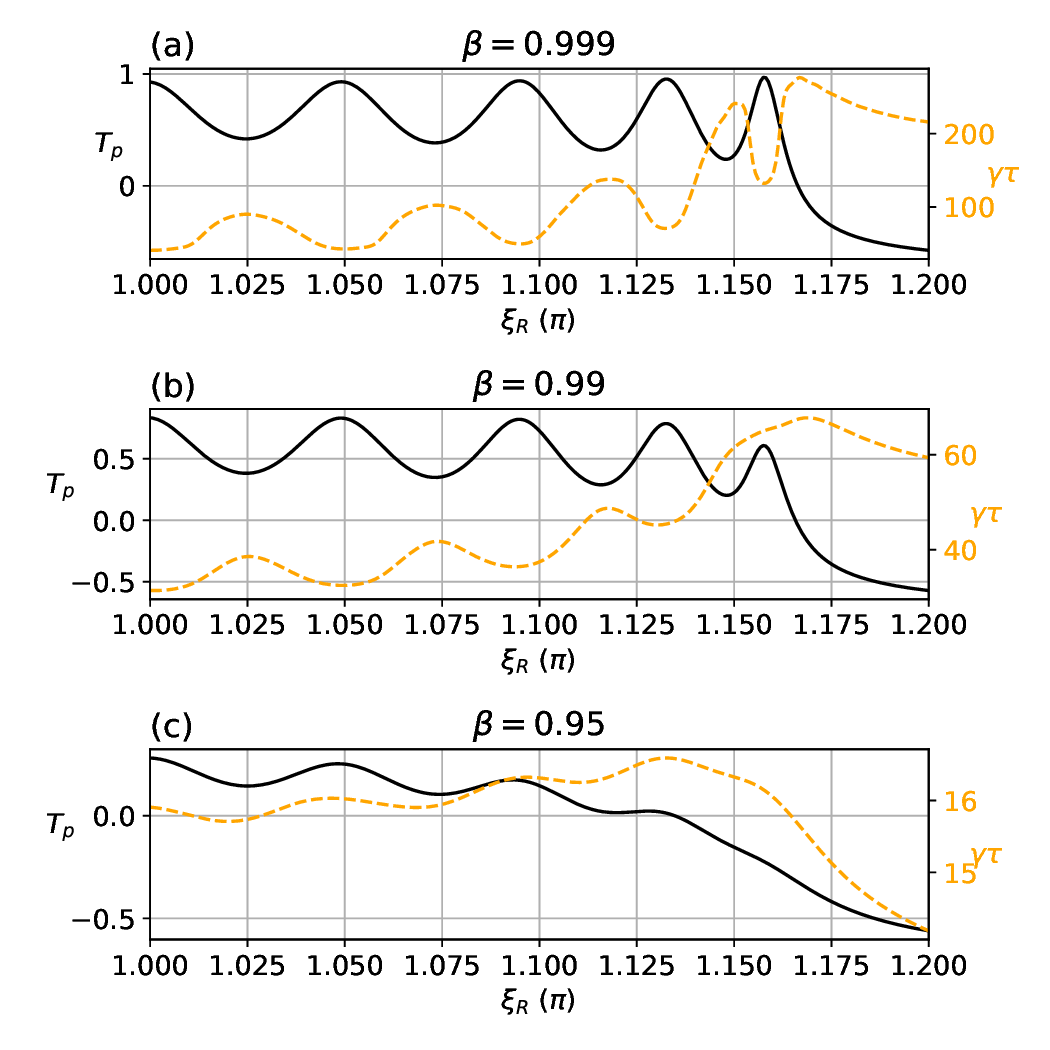}
	\caption{$T_p$ and $\tau$ with respect to $\xi_R$ on the cross-sectional cut (the light green line) in Fig. \ref{Fig.S8}. $T_p$ is plotted in black solid line, and $\tau$ is plotted in orange dashed line. All cases are simulated with $N_L=N_R=10$, and $D=0.5$, with interatomic spacings $\xi_L = 1.8\pi$ and $\xi_D = 1.5\pi$. 
	}\label{Fig.S9}
\end{figure*}

For the resonant case ($\delta=0$), one can get the corrected interaction matrix $\tilde M$, with elements:
\begin{align} 
	\tilde M_{\mu\nu}
	=\begin{cases}
		-\gamma_Le^{ik_s|x_\mu-x_\nu|}, & \mu<\nu\\
		-\frac{\gamma}{2\beta}, & \mu=\nu\\
		-\gamma_Re^{ik_s|x_\mu-x_\nu|}, & \mu>\nu\\
	\end{cases}.
\end{align}

Here, we present some representative results for $\beta=0.999$, $\beta=0.99$ and $\beta=0.95$ corresponding to the coupling strength of a superconducting qubits platform \cite{Mirhosseini2019}, a quantum dots platform \cite{Arcari2014}, and a case where our prediction cannot be applied, respectively. We show the phase diagram of excitation transport $T_p$ with respect to $\xi_L$ and $\xi_R$ for these three cases in Fig. \ref{Fig.S8} with all other parameters being the same. One can observe that the intensity of the fringes is weakened as $\beta$ decreases.

In Fig. \ref{Fig.S9}, we show plots of $T_p$ and $\tau$ on the cross-sectional cut in Fig. \ref{Fig.S8} for $\xi_L=1.8\pi$. One can see that for $\beta=0.999$ and $\beta=0.99$, as shown in Fig. \ref{Fig.S9}(a) and (b), the coincidence of relatively high excitation transport and relatively low evolution time is still valid on these two platforms. However, for $\beta=0.95$, shown in Fig. \ref{Fig.S9}(c), the coincidence is lost. The on-site nonguided decay increases the tendency of excitations to be localized in the driven group, reducing the amount of transport.

\subsection{Robustness Against Disorder}
\begin{figure*}[h]
	\centering
	\includegraphics[width=0.65\textwidth]{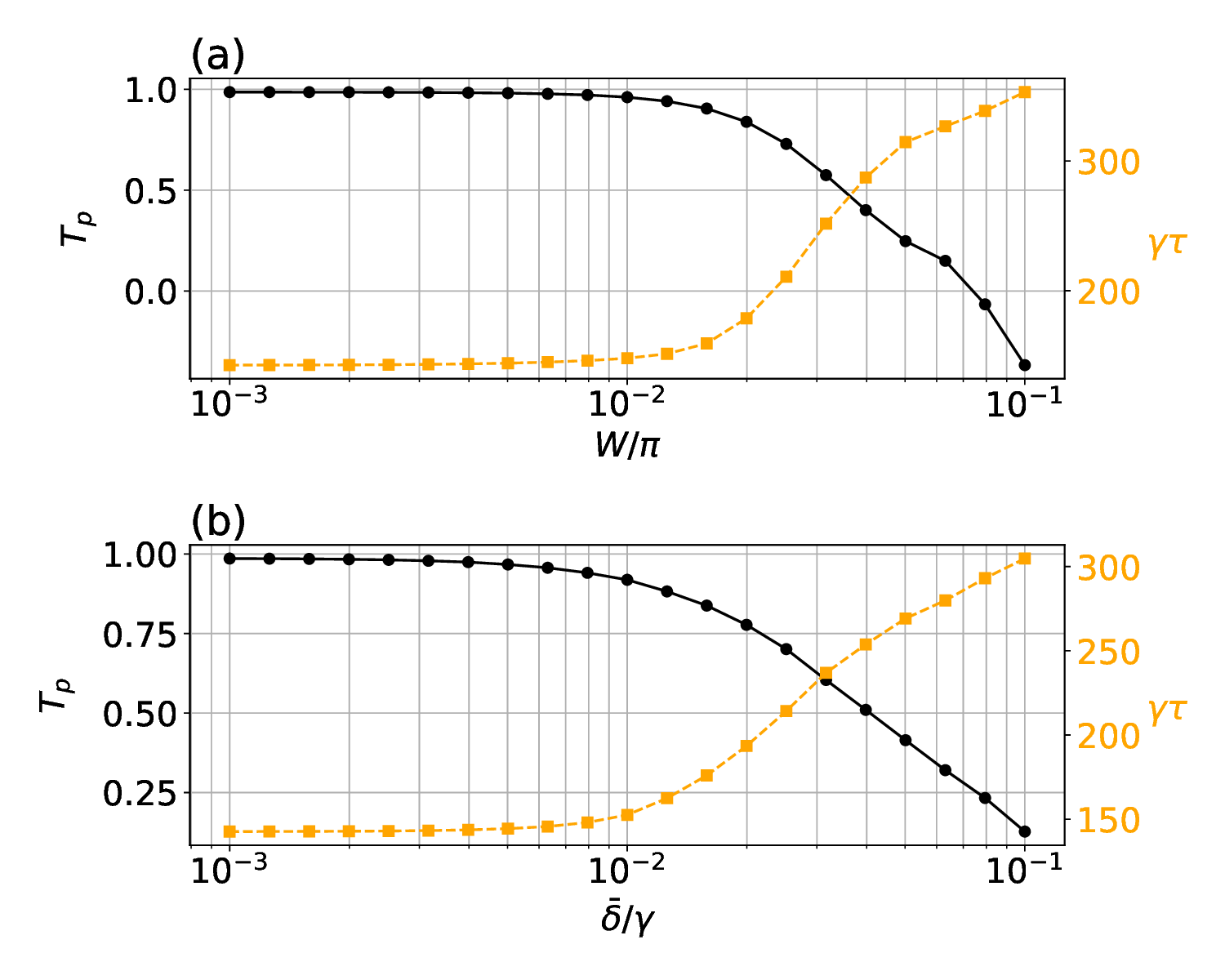}
	\caption{$T_p$ and $\tau$ with respect to (a) the maximum phase disorder $W$, and (b) the maximum detuning disorder $\bar{\delta}$. The parameters are $N=20$, $D=0.5$ with $\xi_L=1.8\pi$, $\xi_D=1.5\pi$, and $\xi_R=1.158$, corresponding to the optimal configuration in Fig. \ref{Fig.3}(a). For each data point, we take the average of 10000 trials to ensure statistical convergence.
	}\label{Fig.S10}
\end{figure*}

Beyond dissipative loss, we also consider structural and spectral imperfections. We model positional disorder by introducing random phase fluctuations $\theta_\mu \in [-W, W]$ to the atomic positions, and inhomogeneous broadening by introducing random detunings $\delta_\mu \in [-\bar{\delta}, \bar{\delta}]$ for each atom.
Fig. \ref{Fig.S10} displays the ensemble-averaged $T_p$ and $\tau$ as a function of the disorder strengths $W$ and $\bar{\delta}$, calculated using the optimal parameters from Fig. \ref{Fig.3}(a). As shown in Fig. \ref{Fig.S10}(a), the transport remains robust against positional disorder, maintaining $>90\%$ efficiency for $W \lesssim 0.02\pi$. Similarly, Fig. \ref{Fig.S10}(b) demonstrates that the system tolerates inhomogeneous broadening up to $\bar{\delta} \approx 0.01\gamma$ before significant degradation occurs. These thresholds indicate that the proposed transport mechanism is feasible within typical experimental linewidths and fabrication tolerances.

\end{document}